\documentclass[12pt]{iopart}

\usepackage{iopams}
\usepackage{graphicx}
\usepackage{hyperref}
\usepackage{multirow}
\usepackage{cancel}
\usepackage{color}

\begin{document}

\title[Understanding partonic energy loss ...]{Understanding partonic energy loss from measured light charged particles and jets in PbPb collisions at LHC energies}

\author{Prashant Shukla$^{1,2 \dagger}$ and Kapil Saraswat$^{3, \ast}$ }

\address{$^{1}$Nuclear Physics Division, Bhabha Atomic Research Center, Mumbai 400085, India.}
\address{$^{2}$Homi Bhabha National Institute, Anushakti Nagar, Mumbai 400094, India.}
\address{$^{3}$Department of Physics, DSB Campus, Kumaun University, Nainital - 263001, INDIA.}
\ead{$^{\dagger}$  pshuklabarc@gmail.com}
\ead{$^{\ast}$     kapilsaraswatbhu@gmail.com}

\vspace{10pt}

\date{\bf{\today}}

\begin{abstract}

  We perform a comprehensive study of partonic energy loss reflected in the
nuclear modification factors of charged particles and jets measured 
in PbPb collisions at $\sqrt{s_{\rm NN}}$ = 2.76 and 5.02 TeV in wide transverse
momentum ($p_{\rm T}$) and centrality range.
The $p_{\rm T}$ distributions in pp collisions are fitted with a modified power law and
the nuclear modification factor in PbPb collisions can be obtained using effective
shift ($\Delta p_{\rm T}$) in the spectrum measured at different centralities.
Driven by physics consideration, the functional form of energy loss given by
$\Delta p_{\rm T}$ can be assumed
as power law with different power indices in three different $p_{\rm T}$ regions.
The power indices and the boundaries of three $p_{\rm T}$ regions are obtained by
fitting the measured nuclear modification factor as a function of $p_{\rm T}$ in all
collision centralities simultaneously. The energy loss in different collisions
centralities are described in terms of fractional power of number of participants.
It is demanded that the power law functions in three $p_T$ regions and their derivatives
are continuous at the $p_{\rm T}$ boundaries. The $\Delta p_{\rm T}$ for light
charged particles is found to increase linearly with $p_{\rm T}$ in low $p_{\rm T}$ region
below $\sim 5-6$ GeV/$c$ and approaches a constant value in high $p_{\rm T}$ region above
$\sim 22-29$ GeV/$c$ with an intermediate power law connecting the two regions.
The method is also used for jets and it is found that for jets, the $\Delta p_{\rm T}$
increases approximately linearly even at very high $p_{\rm T}$. 
\end{abstract}

%
%
%
%
%


\newpage

\section{Introduction}

 The collisions of heavy ions at ultra relativistic energies are performed to create and
study bulk strongly interacting matter at high temperatures. 
The data from RHIC and LHC provide strong evidences of formation of a new state
of matter known as quark gluon
plasma (QGP) in these collisions \cite{Proceedings:2019drx}.

The light charged hadrons and jets 
transverse momentum ($p_{\rm{T}}$) spectra give insight into the particle production
mechanism in pp collisions. The partonic energy energy loss is reflected in these
particles when measured in heavy ion collisions 
due to jet-quenching \cite{Wang:2003aw} which measures the opacity of the medium.
A modified power law distribution \cite{Tsallis:1987eu, Biro:2008hz, Khandai:2013gva} describes
the $p_{\rm{T}}$ spectra of the hadrons in pp collisions in terms of a
power index $n$ which determines  the initial production in partonic
collisions. In Ref.\cite{Saraswat:2017kpg}, the power law function is applied to heavy ion
collisions as well which includes the transverse
flow in low $p_{\rm{T}}$ region and the in-medium energy loss (also in terms of power law)
in high $p_{\rm{T}}$ region.

  The spectra of hadrons are measured in both pp and AA collisions and
nuclear modification factor ($R_{\rm{AA}}$) is obtained.
The energy loss of partons can be connected to horizontal shift in the scaled hadron 
spectra in AA with respect to pp spectra as done by PHENIX measurement \cite{Adler:2006bw}.
Their measurements of neutral pions upto $p_{\rm T} \sim 10$ GeV/$c$ are consistent with the
scenario where the momentum shift $\Delta p_{\rm T}$ is proportional to $p_{\rm T}$.
  In similar approach, the authors in Ref.~\cite{Wang:2008se}, extracted the fractional energy loss
from the measured nuclear modification factor  of hadrons as a function of $p_{\rm{T}}$ below
10 GeV/$c$ in AuAu collisions at $\sqrt{s_{\rm{NN}}}$ = 200 GeV. They also considered that 
the energy loss increases linearly with $p_{\rm T}$.
  In recent PHENIX work \cite{Adare:2015cua},
fractional energy loss was obtained in the hadron spectrum measured upto $p_{\rm T}=20$ GeV/$c$
in heavy ions collisions at RHIC and LHC energy and is not found to be constant.
This means that a constant fractional energy loss (energy loss varying linearly with $p_{\rm T}$)
can be applicable only to low $p_{\rm T}$ RHIC measurements.

 There are many recent studies which use so-called shift formalism to study the energy loss.
The work in Ref.\cite{Spousta:2015fca} is based on 
shift formalism and  describes the transverse momentum ($p_{\rm{T}}$), rapidity ($y$)
and centrality dependences of the measured jet nuclear modification factor ($R_{\rm{AA}}$)
in PbPb collisions at LHC.
They assume that the energy loss is given by a power law in terms of $p_{\rm T}$, the value of power
index is obtained between 0.4 to 0.8 by fitting the $R_{\rm{AA}}$ as a function of $p_{\rm{T}}$
and centralities.
 They also found that the energy loss linearly increases with number of participants. 
Using the same method they study the magnitude and the colour charge dependence of the
energy loss in PbPb collisions at LHC energies using the measured data of the inclusive
jet suppression~\cite{Spousta:2016agr}.
 The authors of the Ref.\cite{Ortiz:2017cul} work on inclusive charged particle spectra
measured in the range ($5 < p_{\rm T} < 20$ GeV/$c$) in heavy ion collisions at RHIC and LHC.
They assume that the energy loss linearly increases with $p_{\rm T}$ and pathlength.



There are detailed calculations of energy loss of partons in the hot medium
[see e.g. Refs.~\cite{Wang:1994fx,Baier:1996kr}.
 Phenomenological models tend to define simple dependence of the radiative energy
loss of the parton on the energy of the parton inside the medium [for a discussions
see Ref.~\cite{Baier:2000mf}]. The energy loss can be characterized in terms of 
coherence length $l_{\rm{coh}}$, which is associated with the formation time of
gluon radiation by a group of scattering centres. If $l_{\rm{coh}}$ is less then 
the mean free path $(\lambda)$ of the parton, the energy loss is proportional to the
energy of the parton.
If $l_{\rm{coh}}$ is greater than $\lambda$ but less than the path length ($L$) of the
parton ($\lambda < l_{\rm{coh}} < L)$, the energy loss is proportional to the square
root of the energy of the parton.
In the complete coherent regime, $l_{\rm{coh}} > L$, the energy loss per unit length
is independent on energy but proportional to the parton path length implying that
$\Delta p_T$ is proportional to square of pathlength.
 There is a nice description of charged particle spectra at RHIC and LHC using such a
prescription by dividing the $p_{\rm T}$ spectra in three regions \cite{De:2011fe, De:2011aa}.
  For low and intermediate energy partons, $\Delta p_T$ is assumed to be linearly
dependent on $L$ \cite{Muller:2002fa}. The work in Ref.~\cite{Betz:2011tu} studies 
the energy loss of jets in terms of exponent of the number of participants.
 It should be remembered that the fragmentation changes the momentum between the partonic
stage (at which energy is lost) and hadron formation.
 There are models which say that softening occurs at fragmentation stage 
due to color dynamics [See e.g. Ref.~\cite{Beraudo:2012bq}].

 In general, one can assume that the energy loss of partons in the hot medium as a function of
parton energy is in the form of power law where the power index ranges from 0 (constant) to
1 (linear). Guided by these considerations, in the present work, the $p_{\rm T}$ loss has been assumed
as power law with different power indices in three different $p_{\rm T}$ regions.
The energy loss in different collisions centralities are described in terms of fractional power
of number of participants.
  The $p_{\rm T}$ distributions in pp collisions are fitted with a modified power law and
$R_{\rm AA}$ in PbPb collisions can be obtained using effective shift ($\Delta p_{\rm T}$) in the
$p_{\rm T}$ spectrum measured at different centralities.
 The power index and the boundaries of three $p_{\rm T}$ regions are obtained by fitting the
measured $R_{\rm{AA}}$ of charged particles and jets in PbPb collisions at
$\sqrt{s_{\rm NN}}$ =  2.76 and 5.02 TeV in large transverse momentum ($p_{\rm T}$) and
centrality range.
The shift $\Delta p_{\rm{T}}$ includes the medium effect, mainly energy loss of parent
quark inside the plasma.
 The shift $\Delta p_T$ can be approximatively understood as the partonic energy loss in the
case of jets while in case of hadrons it is not simple due to complicated correlations.
Often we refer to the shift $\Delta p_{\rm{T}}$ as the energy loss.

\section{Nuclear Modification Factor and Energy Loss}

The nuclear modification factor $R_{\rm{AA}}$ of hadrons is defined as the ratio
of yield of the hadrons in AA collision and the yield in pp collision with a
suitable normalization
\begin{equation}
R_{\rm{AA}} (p_{\rm{T}}, b) = \frac{1}{T_{\rm{AA}}} {\frac{d^2N^{AA}(p_{\rm{T}}, b)}{dp_{\rm{T}}dy}}/
{\frac{d^2\sigma^{pp}(p_{\rm{T}}, b)}{dp_{\rm{T}}dy}}~.
\label{raa_definition}
\end{equation}
Here,  $T_{\rm{AA}}$ is the nuclear overlap function which can be calculated from the
nuclear density distribution. High $p_{\rm T}$ partons traversing the medium loose energy and
cause the suppression of hadrons at high $p_{\rm T}$ indicated by value of $R_{\rm{AA}}$
which is less than one.
 The transverse momentum distribution of hadrons in pp collisions
can be described by the Hagedorn function which is a  QCD-inspired summed power
law \cite{Hagedorn:1983wk} given as
\begin{equation}
\,\,\,\,\,\,\,\,\,\,\,   \frac{d^2\sigma^{\rm{pp}}}{dp_{\rm{T}}dy}
= A_n~2\pi p_{\rm{T}} ~\Bigg(1 + \frac{p_{\rm{T}}}{p_{0}}\Bigg)^{-n}~.
\label{Hag}
\end{equation}
where $n$ is the power and $A_n$ and $p_{0}$ are other parameters which are obtained
by fitting the experimental pp spectrum.
  The yield in the AA collision can be obtained by shifting the spectrum by 
$\Delta p_{\rm T}$ as
\begin{eqnarray}
 \,\,\,\,\,  \frac{1}{T_{\rm{AA}}}\frac{d^{2}N^{\rm{AA}}}{dp_{\rm{T}}dy}
   &  = \frac{d^{2}\sigma^{\rm{pp}}(p'_{\rm{T}} = p_{\rm{T}} +  \Delta p_{\rm{T}})}{dp'_{\rm{T}}dy}
  \frac{dp'_{\rm{T}}}{dp_{\rm{T}}}  \nonumber \\
  & = \frac{d^{2}\sigma^{\rm{pp}}(p'_{\rm{T}})}{dp'_{\rm{T}}dy}
  \Bigg(1 + \frac{d(\Delta p_{\rm{T}})}{dp_{\rm{T}}}\Bigg)~.
\label{shiftRAA}
\end{eqnarray}
The reasoning behind writing Eq.~\ref{shiftRAA} lies in the assumption that particle yield at
a given $p_{\rm{T}}$ in AA collisions would have been equal to the yield in pp collisions
at $p_{\rm{T}} + \Delta p_{\rm{T}}$. The shift $\Delta p_{\rm{T}}$ includes the medium effect,
mainly energy loss of parent quark inside the plasma.

The nuclear modification factor $R_{\rm{AA}}$ can be obtained as 
\begin{eqnarray}
R_{\rm{AA}} = \left(1 + { \Delta p_{\rm{T}} \over p_{0}+p_{\rm{T}} } \right)^{-n} \,\,
\left({p_{\rm{T}} + \Delta p_{\rm{T}} \over  p_{\rm{T}}}\right) \, 
\left(1 + {d(\Delta p_{\rm{T}}) \over dp_{\rm{T}}}\right)
\label{nmf_raa_fitting_function}
\end{eqnarray}
The energy loss given by $p_{T}$ loss, $\Delta p_{T}$ can be extracted by fitting the
experimental data on $R_{\rm AA}$
with Eq.~\ref{nmf_raa_fitting_function}.
The $\Delta p_{\rm T}$ and its derivative will go as input in the above equation
and can be assumed to be in the form of the power law
with different values of power indices in three different $p_{\rm T}$ regions as follows


\begin{eqnarray}
\Delta p_{\rm T}  =   \left\{
\begin{array}{l}
a_1~(p_{\rm T} - C_1)^{\alpha_1} ~~~ {\rm for} ~~~ p_{\rm T} < p_{\rm T_1}~~~, \\
a_2~(p_{\rm T} - C_2)^{\alpha_2} ~~~ {\rm for} ~~~ p_{\rm T_1} \leq p_{\rm T} < p_{\rm T_2}~~,\\
a_3~(p_{\rm T} - C_3)^{\alpha_3} ~~~ {\rm for} ~~~ p_{\rm T} \geq p_{\rm T_2}~.
\end{array}
\right\}
\label{Equation_Two}
\end{eqnarray}

The parameter $a_1$ in our work contains the pathlength dependence. The pathlength $L$
 scales as the square root of number of participants as $\sqrt{N_{\rm part}}$.
 For low and intermediate energy partons, $\Delta p_T$ can be assumed to be
linearly dependent on $L$ \cite{Muller:2002fa}. If the scattering happens
in complete coherent regime where the whole medium acts as one coherent source of radiation, 
the $\Delta p_T$ approaches quadratic dependence on $L$.
 The work in Ref.~\cite{Betz:2011tu} studies 
the energy loss of jets in terms of exponent of the number of participants.
Without complicating the calculations we can assume that
$a_1 = M \, (N_{\rm{part}}/(2A))^\beta$. The exponent $\beta$ is obtained separately
for each dataset. 
 The parameter $M$ relies on the energy density of the medium depending on the
collision energy but has a same value for all centralities.
The boundaries of the $p_{\rm T}$ regions $p_{{\rm{T}}_{1}}$, $p_{{\rm{T}}_{2}}$ and the  power
indices $\alpha_{1}$, $\alpha_{2}$ and $\alpha_{3}$ in the three different regions are
used as free parameters while fitting the $R_{\rm{AA}}$ measured at different centralities
simultaneously.
 The parameter $C_{1}$ is fixed to a suitable value to choose a lower $p_{\rm T}$ cutoff
 and the parameters $C_{2}$, $C_{3}$, $a_2$ and $a_3$ are obtained by assuming the function
 and its derivative to be continuous at boundaries.

Demanding that the function in Eq.~\ref{Equation_Two} to be continuous
at $p_{\rm{T}} = p_{{\rm{T}}_{1}}$ and at $p_{\rm{T}} = p_{{\rm{T}}_{2}}$ we obtain
\begin{equation}
  a_{2} = a_{1}~ \frac{(p_{{\rm{T}}_{1}} - C_{1})^{\alpha_{1}}}{(p_{{\rm{T}}_{1}} - C_{2})^{\alpha_{2}}}~~.
\label{Equation_Three}
\end{equation}

\begin{equation}
  a_{3} = a_{2}~ \frac{(p_{{\rm{T}}_{2}} - C_{2})^{\alpha_{2}}}{(p_{{\rm{T}}_{2}} - C_{3})^{\alpha_{3}}}~~.
\label{Equation_Four}
\end{equation}
 Demanding that at $p_{\rm{T}} = p_{{\rm{T}}_{1}}$, the derivative of Eq.~\ref{Equation_Two} is  continuous.
\begin{equation}
 a_{1}~\alpha_{1}~(p_{{\rm{T}}_{1}} - C_{1})^{(\alpha_{1}-1)} = 
 a_{2}~\alpha_{2}~(p_{{\rm{T}}_{1}} - C_{2})^{(\alpha_{2}-1)}~~,
\label{Equation_Seven}
\end{equation}
Using the value of $a_{2}$ from Eq.~\ref{Equation_Three}
\begin{equation}
\frac{\alpha_{1}}{(p_{{\rm{T}}_{1}} - C_{1})} = \frac{\alpha_{2}}{(p_{{\rm{T}}_{1}} - C_{2})}~~,
\label{Equation_Eight}
\end{equation}

\begin{equation}
C_{2} = p_{{\rm{T}}_{1}} - \frac{\alpha_{2}}{\alpha_{1}}(p_{{\rm{T}}_{1}} -C_{1})~.
\label{Equation_Nine}
\end{equation}
Similarly, demanding the derivative to be continuous at
$p_{\rm{T}} = p_{{\rm{T}}_{2}}$, we get $C_{3}$ 
\begin{equation}
C_{3} = p_{{\rm{T}}_{2}} - \frac{\alpha_{3}}{\alpha_{2}}(p_{{\rm{T}}_{2}} -C_{2})~.
\label{Equation_Ten}
\end{equation}
In case of jets we consider only one region as the data starts from very high $p_T$
above 40 GeV/$c$.

\section{Results and Discussions}

\begin{figure}[htp]
\centering
\includegraphics[width=0.60\linewidth]{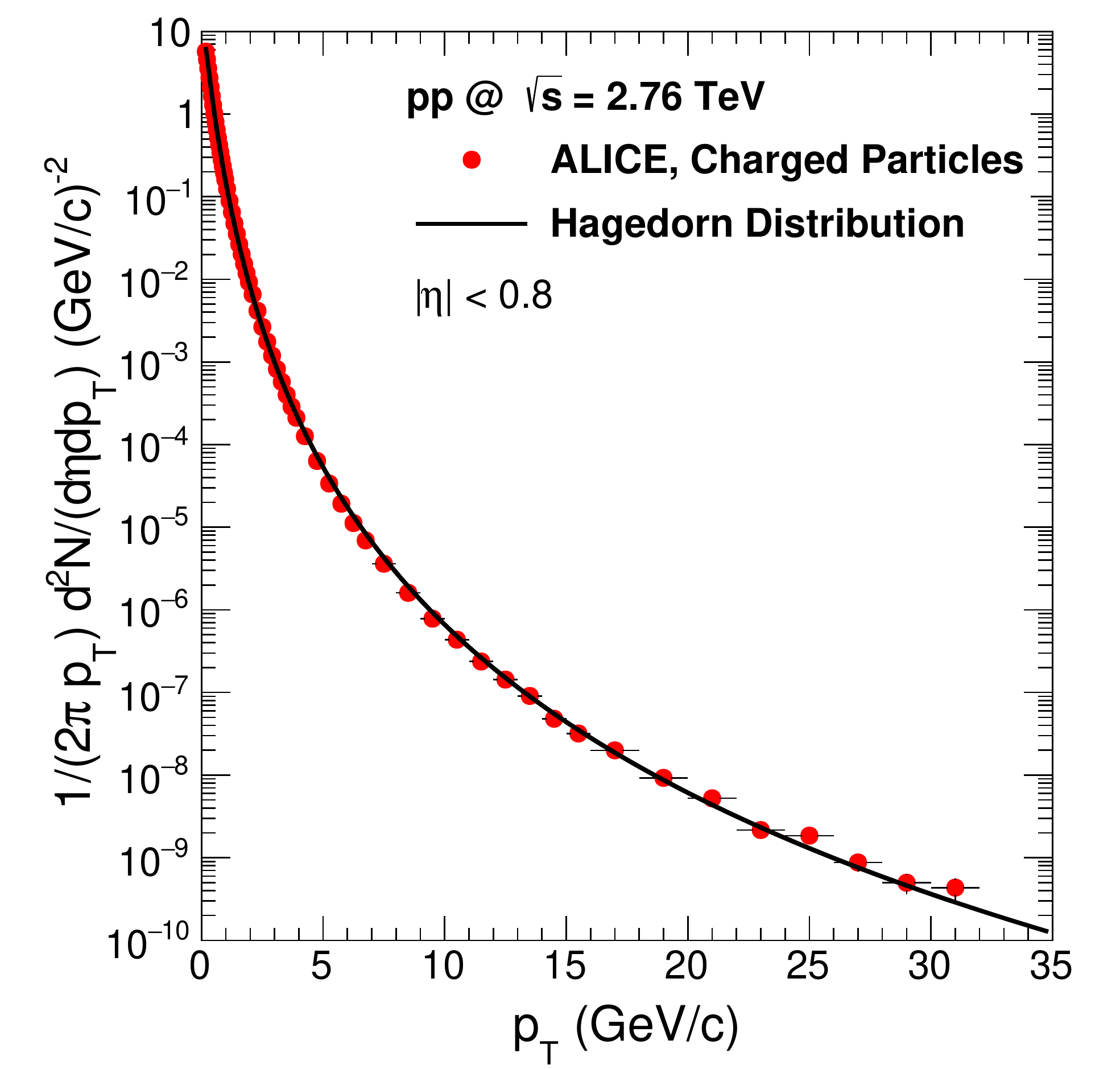}
\caption{The invariant yields of the charged particles as a function of transverse momentum 
$p_{\rm{T}}$ for pp collision at $\sqrt{s}$ = 2.76 TeV measured by the ALICE experiment
 \cite{Abelev:2013ala}. The solid curve is the fitted Hagedorn function.}
\label{Figure1_charged_particles_pT_ALICE_spectra_Tsallis_fit_pp_276TeV}
\end{figure}

Figure~\ref{Figure1_charged_particles_pT_ALICE_spectra_Tsallis_fit_pp_276TeV} shows the
invariant yields of the charged particles as a function of the transverse momentum $p_{\rm{T}}$
for pp collisions at $\sqrt{s}$ = 2.76 TeV measured by the ALICE
experiment \cite{Abelev:2013ala}. The solid curve is the Hagedorn distribution fitted
to the $p_{\rm{T}}$ spectra with the parameters given in 
Table~\ref{table0_charged_particles_jet_pT_spectra_tsallis_fitting_parameters_276_502_TeV}. 

\begin{table}[ht]
  \caption{Parameters for the Hagedorn function obtained by fitting the
transverse momentum spectra of charged particles and jets measured in pp 
collisions at $\sqrt{s_{\rm{NN}}}$ = 2.76 and 5.02 TeV.}
\begin{center}
\scalebox{0.8}{
\begin{tabular}{| c || c | c | c | c |} 
\hline
Parameters  & \multicolumn{2}{c|}{Charged particles}
                         & \multicolumn{2}{c|}{Jets } \\ \hline
            & $\sqrt{s_{\rm{NN}}}$ = 2.76 TeV & $\sqrt{s_{\rm{NN}}}$ = 5.02 TeV 
                        & $\sqrt{s_{\rm{NN}}}$ = 2.76 TeV & $\sqrt{s_{\rm{NN}}}$ = 5.02 TeV  \\  \hline \hline
  $n$                & 7.26  $\pm$ 0.08 & 6.70  $\pm$ 0.14     & 8.21  $\pm$ 1.55   & 7.90 $\pm$ 0.50  \\\hline
  $p_{0}$ (GeV/$c$)  & 1.02  $\pm$ 0.04  & 0.86  $\pm$ 0.16     & 18.23  $\pm$ 1.69  & 19.21 $\pm$ 3.20 \\\hline
$\chi^{2}/\rm{NDF}$ & 0.15 & 0.06 & 0.23 & 0.95  \\\hline  
\end{tabular}}
\end{center}
\label{table0_charged_particles_jet_pT_spectra_tsallis_fitting_parameters_276_502_TeV}
\end{table}

\begin{figure}[htp]
\centering
\includegraphics[width=0.9\linewidth]{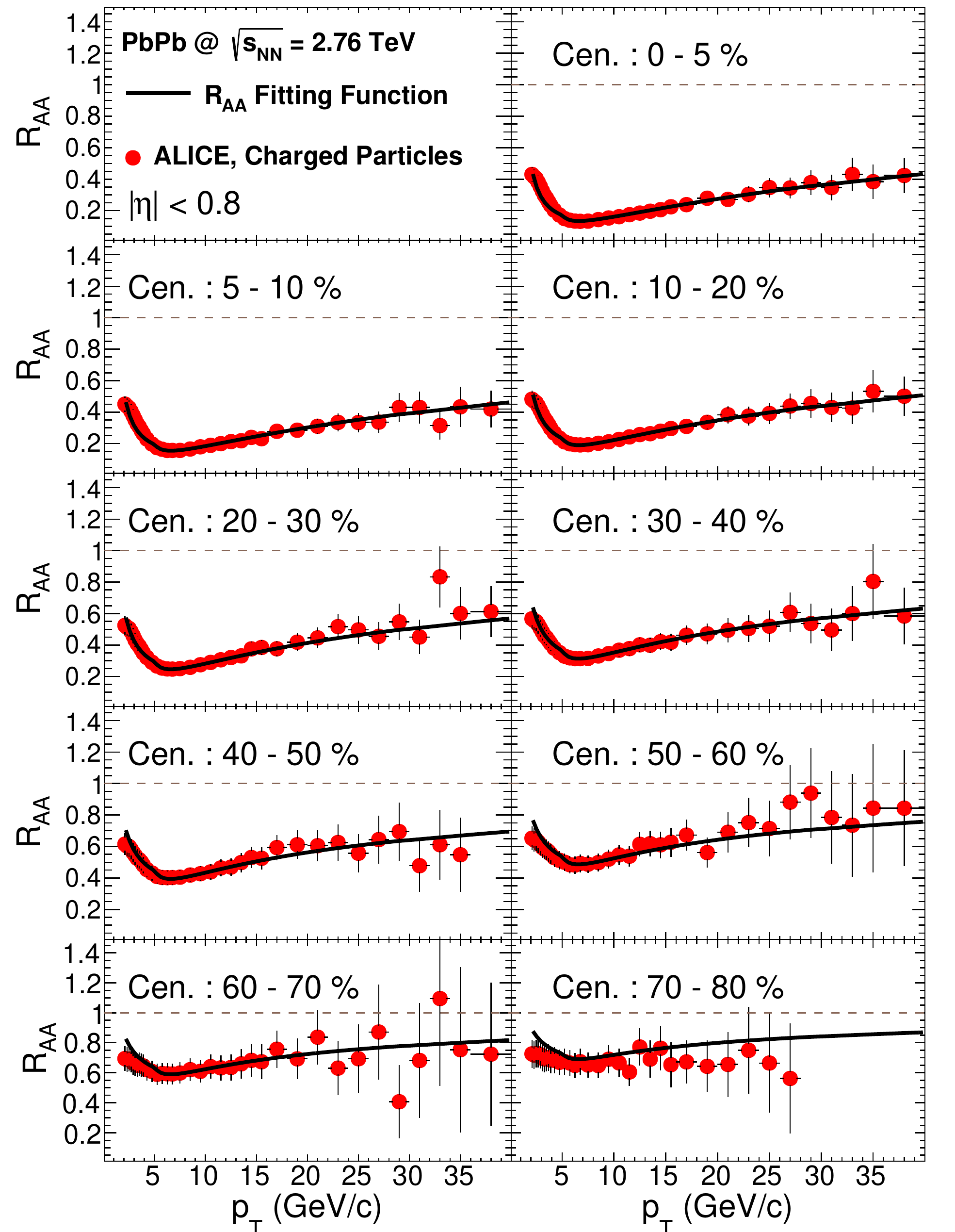}
\caption{The nuclear modification factor $R_{\rm{AA}}$  of the charged particles as a 
function of transverse momentum $p_{\rm{T}}$ for various centrality classes in PbPb 
collisions at $\sqrt{s_{\rm{NN}}}$ = 2.76 TeV measured by the ALICE experiment \cite{Abelev:2012hxa}.
 The solid curves are the $R_{\rm{AA}}$ fitting function
(Eq.~\ref{nmf_raa_fitting_function}).} 
\label{Figure2_charged_particles_RAA_ALICE_spectra_com_fit_PbPb_276TeV}
\end{figure}

Figure~\ref{Figure2_charged_particles_RAA_ALICE_spectra_com_fit_PbPb_276TeV} shows the nuclear
modification factor $R_{\rm{AA}}$ of the charged particles as a function of the transverse
momentum $p_{\rm{T}}$ for different centrality classes in PbPb collisions at $\sqrt{s_{\rm{NN}}}$
= 2.76 TeV measured by the ALICE experiment \cite{Abelev:2012hxa}. The solid lines are the
function given by Eq.~\ref{nmf_raa_fitting_function}. The modeling of centrality dependence using
$N_{\rm part}^\beta$ with $\beta=0.58$ gives a very good description of the data. 
 The extracted parameters of the shift $\Delta p_{\rm{T}}$ 
obtained by fitting the $R_{\rm{AA}}$ measured in different centrality classes of PbPb
collisions at $\sqrt{s_{\rm{NN}}}$ = 2.76 TeV are given in
Table~\ref{table1_charged_particles_raa_fitting_parameter_276_502_TeV}
along with value of $\chi^{2}/\rm{NDF}$. It shows that the $\Delta p_{\rm{T}}$ increases
almost linearly ($p_{\rm{T}}^{0.97}$) upto $p_{\rm{T}} \simeq$  5 GeV/$c$ in confirmation with
earlier studies. After that it increases slowly with power $\alpha=0.224$ upto
a $p_{\rm{T}}$ value 29 GeV/$c$ and then
becomes constant for higher values of $p_{\rm{T}}$.

\begin{table}[ht]
  \caption[]{The extracted parameters of the shift $\Delta p_{\rm{T}}$  obtained by fitting the charged
    particles $R_{\rm{AA}}$
measured in different centrality classes of PbPb collisions at $\sqrt{s_{\rm{NN}}}$ = 2.76 and 5.02 TeV.}
\label{table1_charged_particles_raa_fitting_parameter_276_502_TeV}
\begin{center}
\begin{tabular}{| c || c | c |} \hline
  ~ Parameters & $\sqrt{s_{\rm{NN}}}$ = 2.76 TeV   & $\sqrt{s_{\rm{NN}}}$ = 5.02 TeV  \\ \hline\hline
~ $M$                        &  0.75  $\pm$  0.02   &   0.80   $\pm$ 0.038    \\ \hline      
~ $p_{{\rm{T}}_{1}}$ (GeV/$c$)     &  5.03  $\pm$  0.15    &   5.10  $\pm$ 0.22     \\ \hline   
~ $p_{{\rm{T}}_{2}}$ (GeV/$c$)     &  29.0   $\pm$  0.1    &   22.2 $\pm$ 4.1      \\ \hline   
~ $C_{1}$         (GeV/$c$)     &   1.0 (fixed)            &   1.0                   \\ \hline  
~ $\alpha_{1}$                &   0.97  $\pm$  0.02    &   0.95   $\pm$ 0.04     \\ \hline  
~ $\alpha_{2}$                &   0.22  $\pm$  0.02    &   0.22  $\pm$ 0.03     \\ \hline    
~ $\alpha_{3}$                &   0.05   $\pm$  0.13     &   0.05   $\pm$ 0.10     \\ \hline     
~ $\frac{\chi^{2}}{\rm{NDF}}$ &   0.35                  &   0.38                  \\ \hline 
\end{tabular}
\end{center}
\end{table}

\begin{figure}[htp]
\centering
\includegraphics[width=0.60\linewidth]{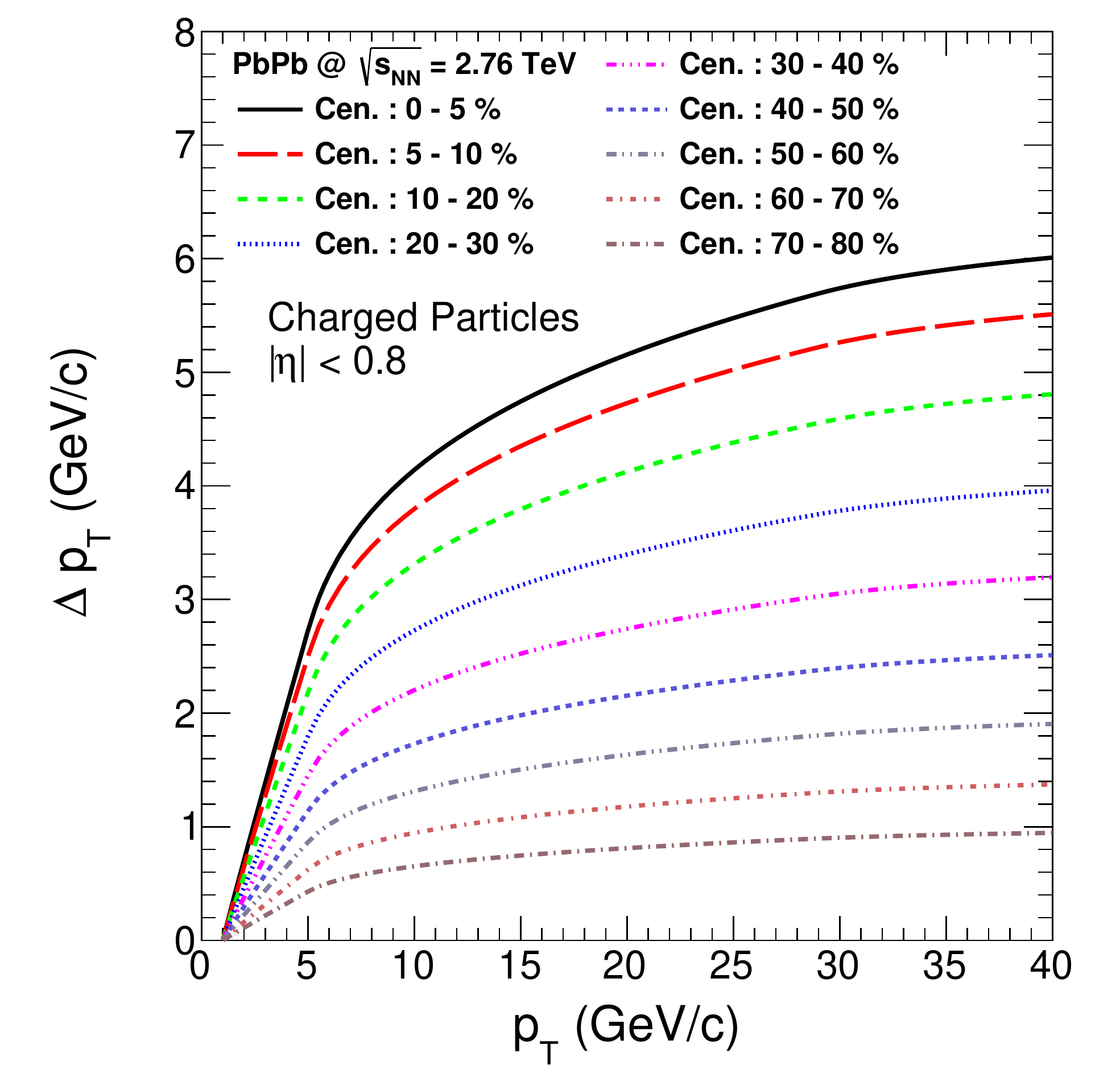}
\caption{The energy loss $\Delta p_{\rm{T}}$ of the charged particles as a function of 
transverse momentum $p_{\rm{T}}$ in PbPb collision at $\sqrt{s_{\rm{NN}}}$ = 2.76 TeV for 
different centrality classes.}
\label{Figure3_charged_particles_com_fit_Del_pT_PbPb_276TeV}
\end{figure}

 Figure~\ref{Figure3_charged_particles_com_fit_Del_pT_PbPb_276TeV} shows the energy loss
$\Delta p_{\rm{T}}$ of the charged particles as a function of the transverse momentum
$p_{\rm{T}}$ for different centrality classes in PbPb collision at $\sqrt{s_{\rm{NN}}}$
= 2.76 TeV. The $\Delta p_{\rm{T}}$ is obtained from Eq.~\ref{Equation_Two}
with parameters given in 
Table~\ref{table1_charged_particles_raa_fitting_parameter_276_502_TeV}.
 The $\Delta p_{\rm{T}}$ increases from peripheral to the most
 central collision regions as per $N_{\rm part}^{0.58}$.
  The figure shows that the $\Delta p_{\rm{T}}$ increases 
almost linearly upto $p_{\rm{T}} \sim 5$ GeV/$c$. After that it
increases slowly upto a $p_{\rm{T}}$ value 29 GeV/$c$ and then
becomes constant for higher values of $p_{\rm{T}}$.

\begin{figure}[htp]
\centering
\includegraphics[width=0.60\linewidth]{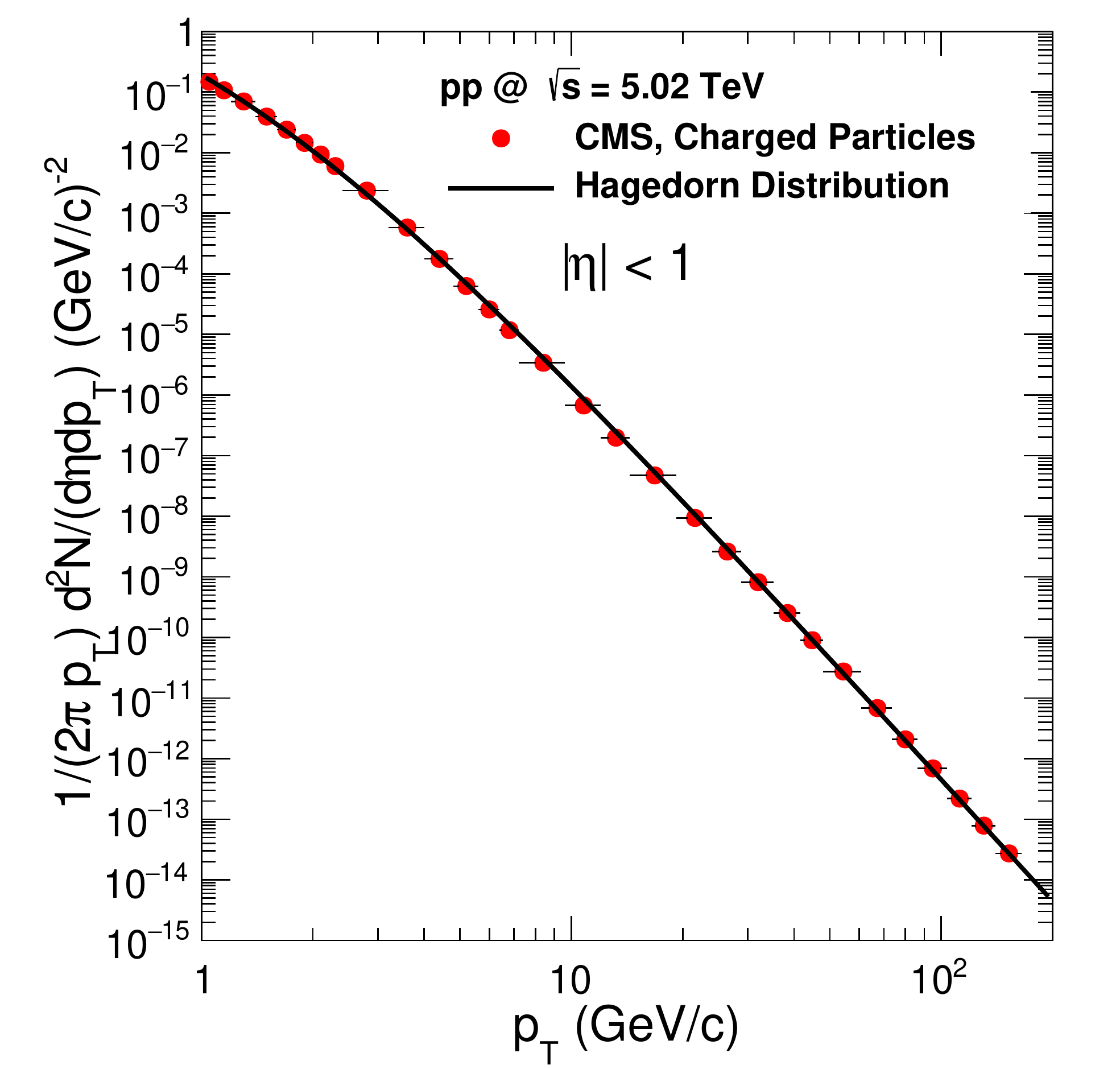}
\caption{The invariant yields of the charged particles as a function of transverse momentum 
$p_{\rm{T}}$ for pp collision at $\sqrt{s}$ = 5.02 TeV measured by the CMS experiment 
\cite{Khachatryan:2016odn}. The solid curve is the fitted Hagedorn function.}
\label{Figure4_charged_particles_cms_pT_spectra_pp_502TeV}
\end{figure}

 Figure~\ref{Figure4_charged_particles_cms_pT_spectra_pp_502TeV} shows the invariant
yields of the charged particles as a function of the transverse momentum $p_{\rm{T}}$
for pp collisions at $\sqrt{s}$ = 5.02 TeV measured by the CMS experiment
\cite{Khachatryan:2016odn}. The solid lines are the Hagedorn function fitted to
the measured $p_{\rm{T}}$ spectra the parameters of which are given in 
Table~\ref{table0_charged_particles_jet_pT_spectra_tsallis_fitting_parameters_276_502_TeV}.

\begin{figure}[htp]
\centering
\includegraphics[width=0.85\linewidth]{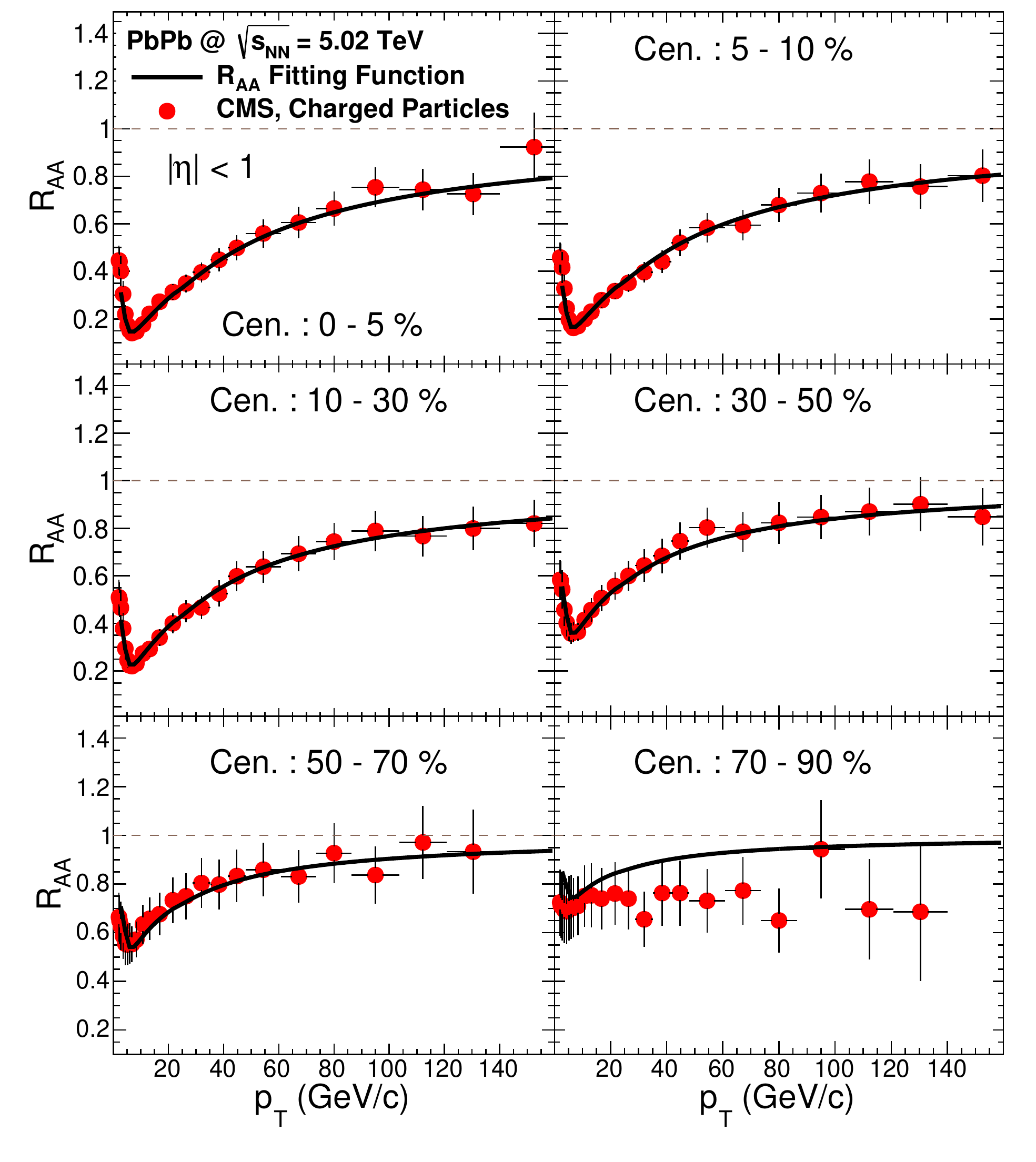}
\caption{The nuclear modification factor $R_{\rm{AA}}$  of the charged particles as a function 
of transverse momentum $p_{\rm{T}}$ for various centrality classes in PbPb collisions at 
$\sqrt{s_{\rm{NN}}}$ = 5.02 TeV measured by the CMS  experiment \cite{Khachatryan:2016odn}. The
solid lines are the $R_{\rm{AA}}$ fitting function (Eq.~\ref{nmf_raa_fitting_function}).}
\label{Figure5_charged_particles_cms_RAA_spectra_com_fit_PbPb_502TeV}
\end{figure}

Figure~\ref{Figure5_charged_particles_cms_RAA_spectra_com_fit_PbPb_502TeV} shows the
nuclear modification factor $R_{\rm{AA}}$ of the charged particles as a function of the
transverse momentum $p_{\rm{T}}$ for different centrality classes in PbPb collisions at
$\sqrt{s_{\rm{NN}}}$ = 5.02 TeV measured by the CMS experiment \cite{Khachatryan:2016odn}.
The solid curves are the $R_{\rm{AA}}$ fitting function (Eq.~\ref{nmf_raa_fitting_function}).
 Here also the modeling of centrality dependence using 
$N_{\rm part}^\beta$ with $\beta=0.58$ gives a good description of the data. 
 The extracted parameters of the shift $\Delta p_{\rm{T}}$ 
obtained by fitting the $R_{\rm{AA}}$ measured in different centrality classes of PbPb
collisions at $\sqrt{s_{\rm{NN}}}$ = 5.02 TeV are given in
Table~\ref{table1_charged_particles_raa_fitting_parameter_276_502_TeV}
along with the value of $\chi^{2}/\rm{NDF}$. It shows that the $\Delta p_{\rm{T}}$ increases
almost linearly ($p_{\rm{T}}^{0.96}$) similar to the case at 2.76 TeV for $p_{\rm T}$ upto 5.1 GeV/$c$.
After that it increases slowly with power $\alpha=0.22$ upto a $p_{\rm{T}}$ value 22.2 GeV/$c$ and then
becomes constant for higher values of $p_{\rm{T}}$ right upto 160 GeV/$c$.

\begin{figure}[htp]
\centering
\includegraphics[width=0.60\linewidth]{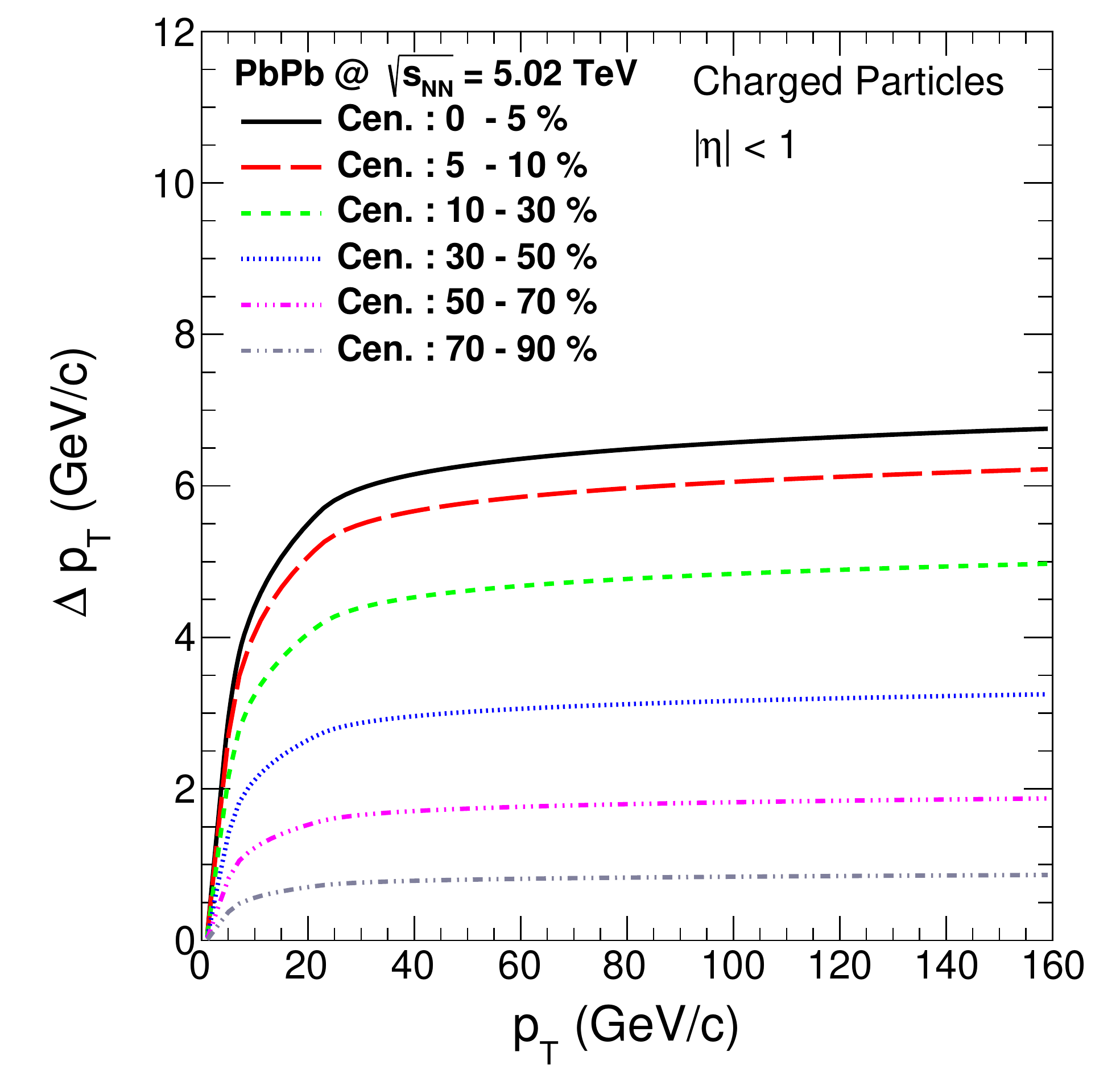}
\caption{The energy loss $\Delta p_{\rm{T}}$ of the charged particles as a function of 
transverse momentum $p_{\rm{T}}$ in PbPb collision at $\sqrt{s_{\rm{NN}}}$ = 5.02 TeV for 
different centrality classes.}
\label{Figure6_charged_particles_com_fit_Del_pT_PbPb_502TeV}
\end{figure}

Figure~\ref{Figure6_charged_particles_com_fit_Del_pT_PbPb_502TeV} shows the energy loss
$\Delta p_{\rm{T}}$ of the charged particles as a function of the transverse momentum
$p_{\rm{T}}$ for different centrality classes in PbPb collision at $\sqrt{s_{\rm{NN}}}$ =
5.02 TeV. The $\Delta p_{\rm{T}}$ is obtained from Eq.~\ref{Equation_Two} with the
parameters given in 
Table~\ref{table1_charged_particles_raa_fitting_parameter_276_502_TeV}.
The $\Delta p_{\rm{T}}$ becomes constant for $p_{\rm{T}}$ in the range 22 GeV/$c$ to 160 GeV/c
and increases as the collisions become more central.

\begin{figure}[htp]
\centering
\includegraphics[width=0.60\linewidth]{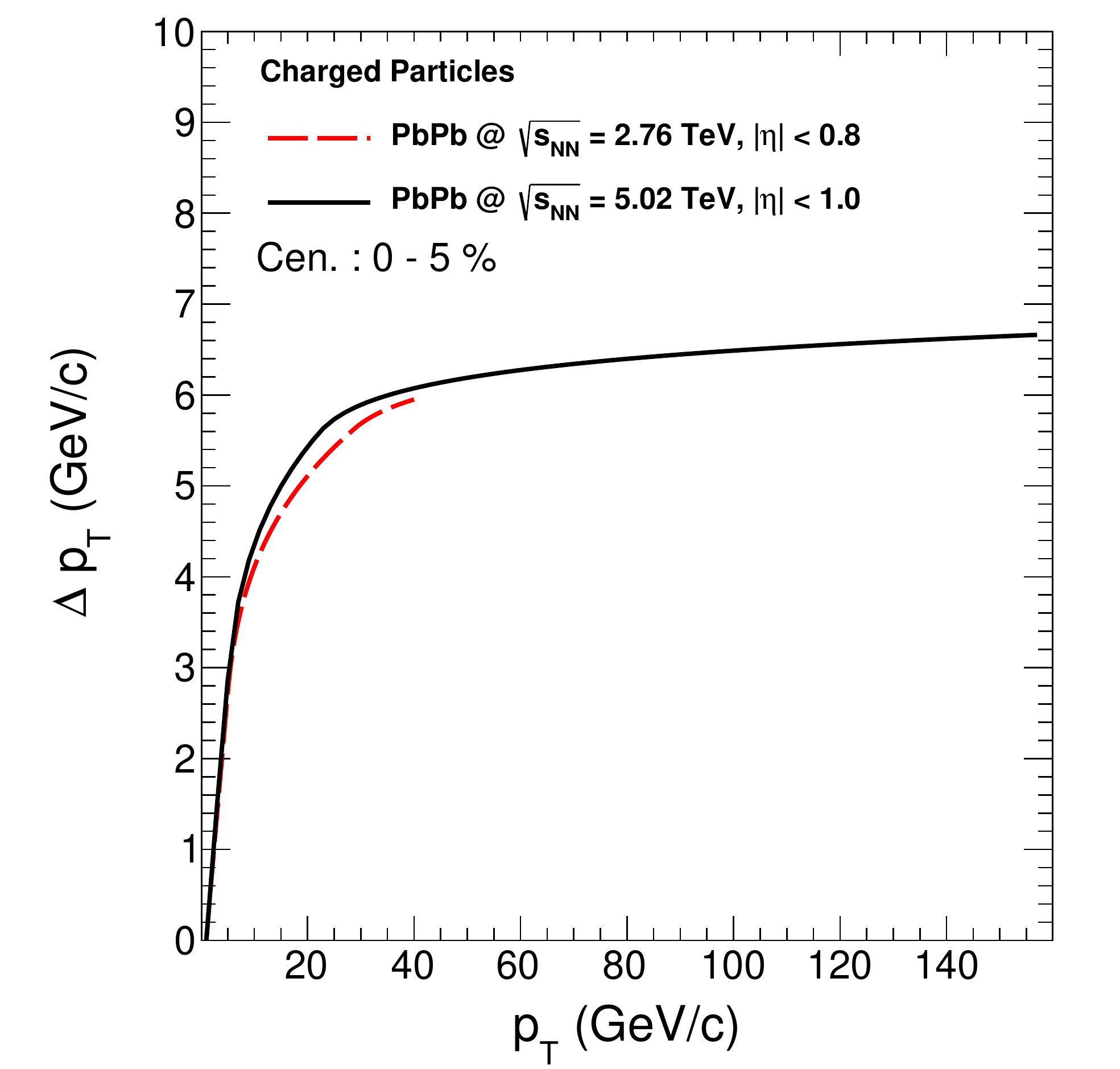}
\caption{The energy loss $\Delta p_{\rm{T}}$ of the charged particles as a function 
of transverse momentum $p_{\rm{T}}$ in PbPb collision at $\sqrt{s_{\rm{NN}}}$ = 2.76 
and 5.02 TeV for 0 - 5 $\%$ centrality.}
\label{Figure7_charged_particles_Del_pT_cen_0_5_PbPb_276_502TeV}
\end{figure}

Figure~\ref{Figure7_charged_particles_Del_pT_cen_0_5_PbPb_276_502TeV} shows the comparison of energy
loss $\Delta p_{\rm{T}}$ of the charged particles as a function of the transverse momentum
$p_{\rm{T}}$ for 0 - 5 $\%$ centrality class in PbPb collision at $\sqrt{s_{\rm{NN}}}$ =
2.76 and at 5.02 TeV. The $\Delta p_{\rm{T}}$ at 5.02 TeV is similar but slightly more than
that at 2.76 TeV.

\begin{figure}[htp]
\centering
\includegraphics[width=0.60\linewidth]{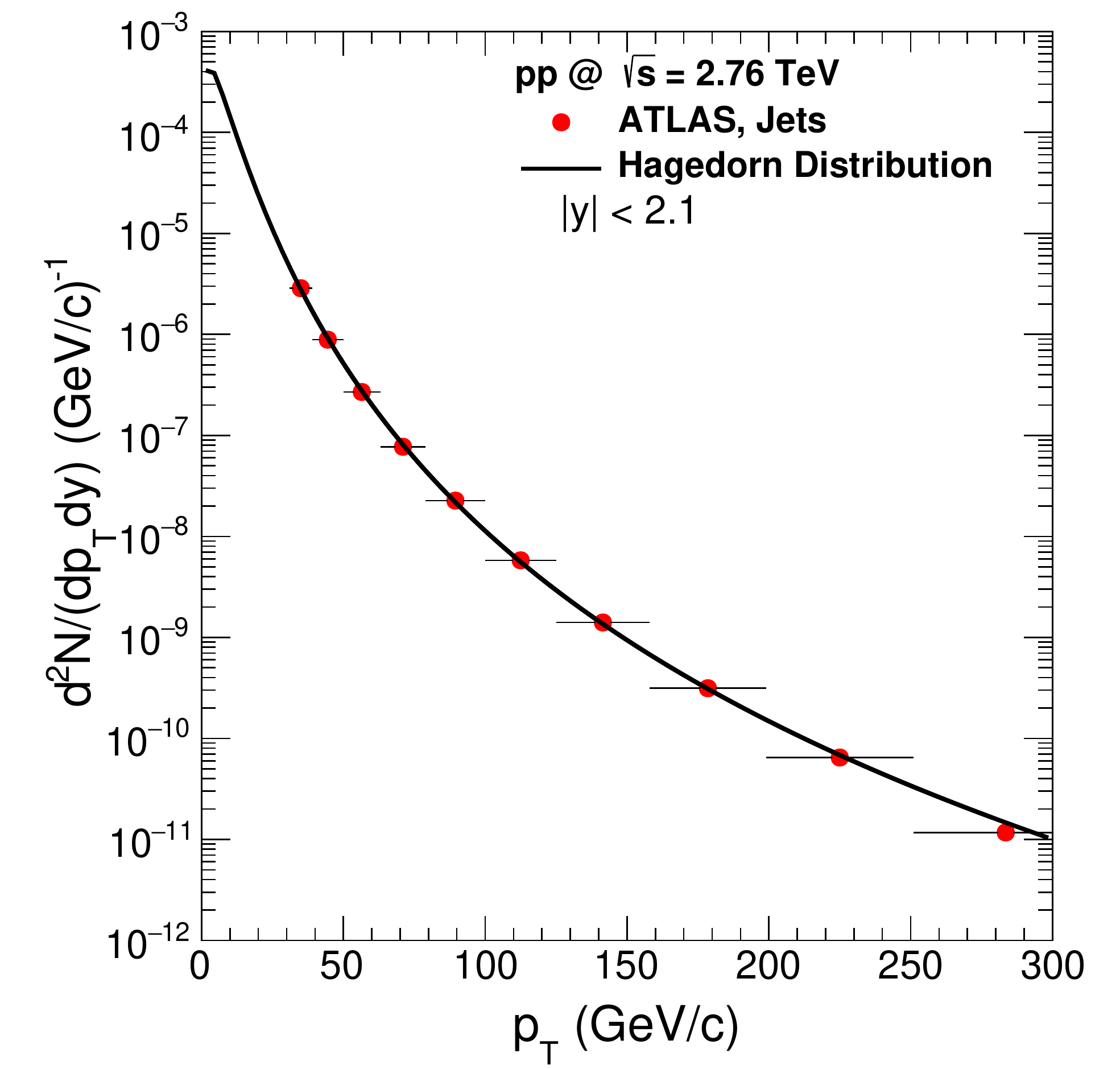}
\caption{The yields of the jets as a function of transverse momentum 
$p_{\rm{T}}$ for pp collision at $\sqrt{s}$ = 2.76 TeV measured by the ATLAS experiment 
  \cite{Aad:2014bxa}. The solid curve is the fitted Hagedorn distribution.}
\label{Figure8_jet_atlas_pT_spectra_pp_276TeV}
\end{figure}

Figure~\ref{Figure8_jet_atlas_pT_spectra_pp_276TeV} shows the yields of the
jets as a function of the transverse momentum $p_{\rm{T}}$ for pp collisions at $\sqrt{s}$
= 2.76 TeV measured by the ATLAS experiment~\cite{Aad:2014bxa}. The solid curve is the
Hagedorn distribution fitted to the $p_{\rm{T}}$ spectra, the parameters of which
are given in 
Table~\ref{table0_charged_particles_jet_pT_spectra_tsallis_fitting_parameters_276_502_TeV}.

\begin{figure}[htp]
\centering
\includegraphics[width=0.85\linewidth]{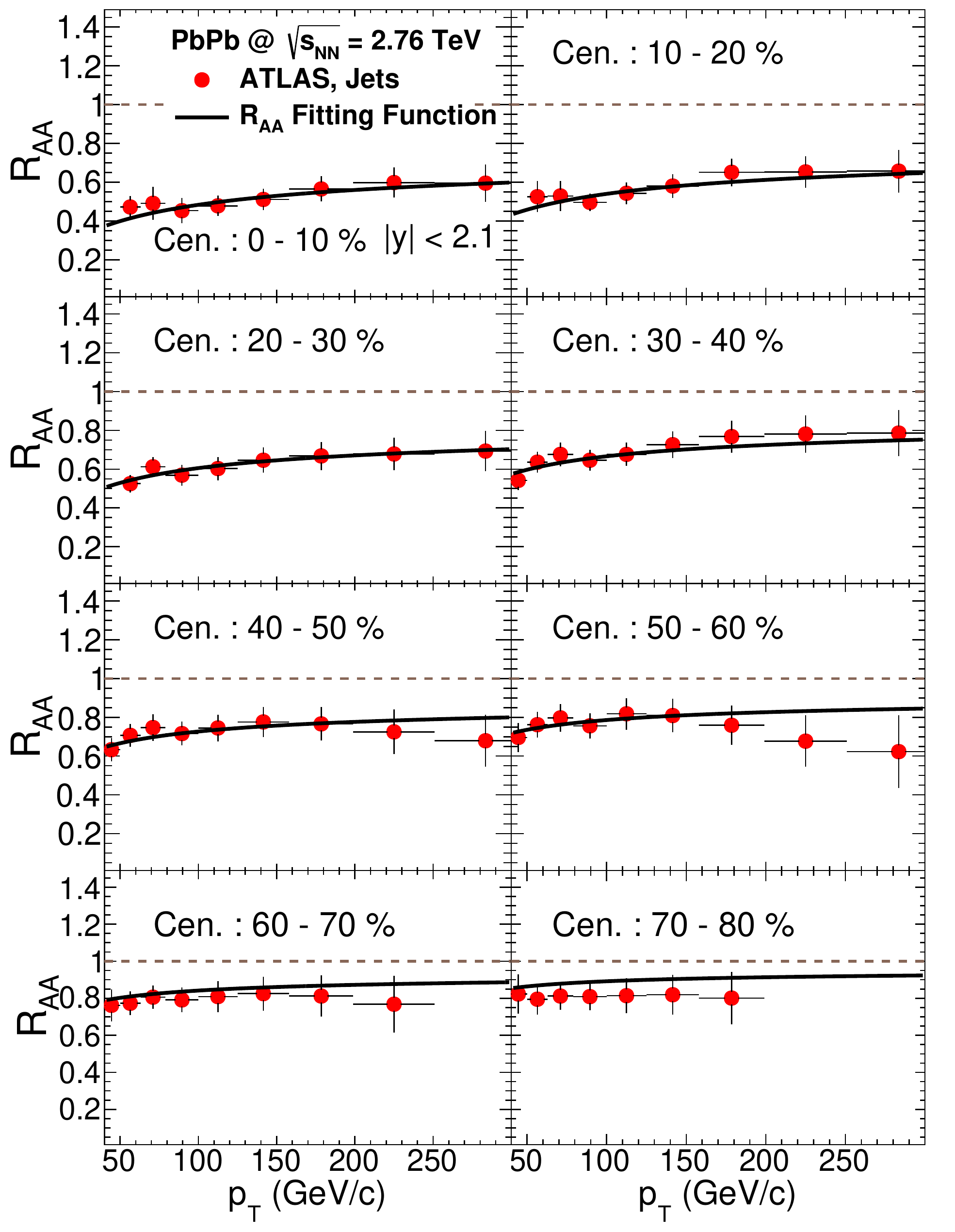}
\caption{The nuclear modification factor $R_{\rm{AA}}$  of jets as a function of
transverse momentum $p_{\rm{T}}$ for various centrality classes in PbPb collisions at 
$\sqrt{s_{\rm{NN}}}$ = 2.76 TeV measured by the ATLAS experiment \cite{Aad:2014bxa}. The solid
curves are the $R_{\rm{AA}}$ fitting function given by Eq.~\ref{nmf_raa_fitting_function}.}
\label{Figure9_Jet_particles_cms_RAA_spectra_com_fit_PbPb_276TeV}
\end{figure}

Figure~\ref{Figure9_Jet_particles_cms_RAA_spectra_com_fit_PbPb_276TeV} shows the nuclear
modification factor $R_{\rm{AA}}$ of the jets as a function of the transverse
momentum $p_{\rm{T}}$ for different centrality classes in PbPb collisions at $\sqrt{s_{\rm{NN}}}$
= 2.76 TeV measured by the ATLAS experiment \cite{Aad:2014bxa}. 
The solid curves are the $R_{\rm{AA}}$ fitting function (Eq.~\ref{nmf_raa_fitting_function}).
 Here the modeling of centrality dependence using $N_{\rm part}^\beta$ with $\beta=0.60$
gives a good description of the data. 
  The extracted parameters of the energy loss
obtained by fitting the $R_{\rm{AA}}$ measured in different centrality classes of PbPb
collisions at $\sqrt{s_{\rm{NN}}}$ = 2.76 TeV are given in
Table~\ref{Table2_Jet_raa_fitting_parameter_PbPb_276_502_TeV}
along with the value of $\chi^{2}/\rm{NDF}$.
It shows that the $\Delta p_{\rm{T}}$ increases as $p_{\rm{T}}^{0.76}$
at all the values of $p_{\rm T}$ measured for jets.

\begin{table}[ht]
  \caption[]{The extracted parameters of the shift $\Delta p_{\rm{T}}$  obtained by fitting the jet $R_{\rm{AA}}$
    measured in different centrality classes of PbPb collisions at $\sqrt{s_{\rm{NN}}}$ = 2.76 and
    5.02 TeV.}
\label{Table2_Jet_raa_fitting_parameter_PbPb_276_502_TeV}
\begin{center}
\begin{tabular}{| c || c | c |} \hline
~ Parameters~  & $\sqrt{s_{\rm{NN}}}$ = 2.76 TeV  & $\sqrt{s_{\rm{NN}}}$ = 5.02 TeV \\ \hline\hline
~ $M$          &  0.33   $\pm$  0.1     &  0.40   $\pm$  0.12     \\ \hline 
~ $C$  (GeV/$c$) &  -55.1  $\pm$ 22.7      &  -119   $\pm$  15        \\ \hline 
~ $\alpha$     &  0.76   $\pm$  0.08    &  0.72  $\pm$  0.01     \\ \hline 
~ $\frac{\chi^{2}}{\rm{NDF}}$  &  0.30    &  0.25                   \\ \hline 
\end{tabular}
\end{center}
\end{table}

\begin{figure}[htp]
\centering
\includegraphics[width=0.60\linewidth]{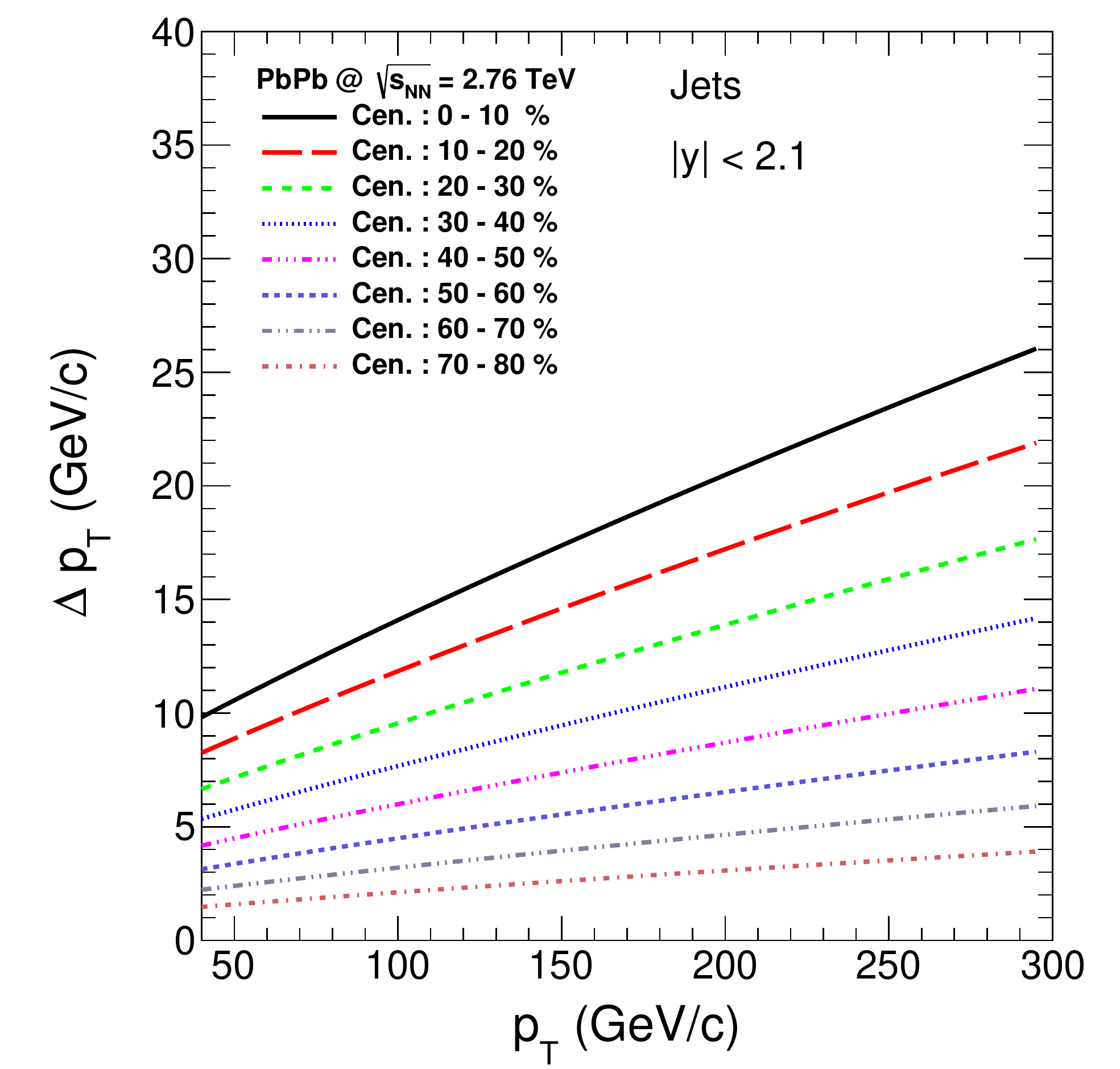}
\caption{The shift  $\Delta p_{\rm{T}}$ of the jets as a function of transverse 
momentum $p_{\rm{T}}$ in PbPb collision at $\sqrt{s_{\rm{NN}}}$ = 2.76 TeV for different 
centrality classes.}
\label{Figure10_Jet_particles_Del_pT_PbPb_276TeV}
\end{figure}

Figure~\ref{Figure10_Jet_particles_Del_pT_PbPb_276TeV} shows the shift $\Delta p_{\rm{T}}$
of the jets as a function of the transverse momentum $p_{\rm{T}}$
for different centrality classes in PbPb collision at $\sqrt{s_{\rm{NN}}}$ = 2.76 TeV.
The $\Delta p_{\rm{T}}$ is obtained from Eq.~\ref{Equation_Two} using the 
parameters given in Table~\ref{Table2_Jet_raa_fitting_parameter_PbPb_276_502_TeV}.
The $\Delta p_{\rm{T}}$ increases from 
peripheral to the most central collision regions.
 The figure shows that the $\Delta p_{\rm{T}}$ increases almost linearly 
at all the values of $p_{\rm T}$ measured for jets.

\begin{figure}[htp]
\centering
\includegraphics[width=0.60\linewidth]{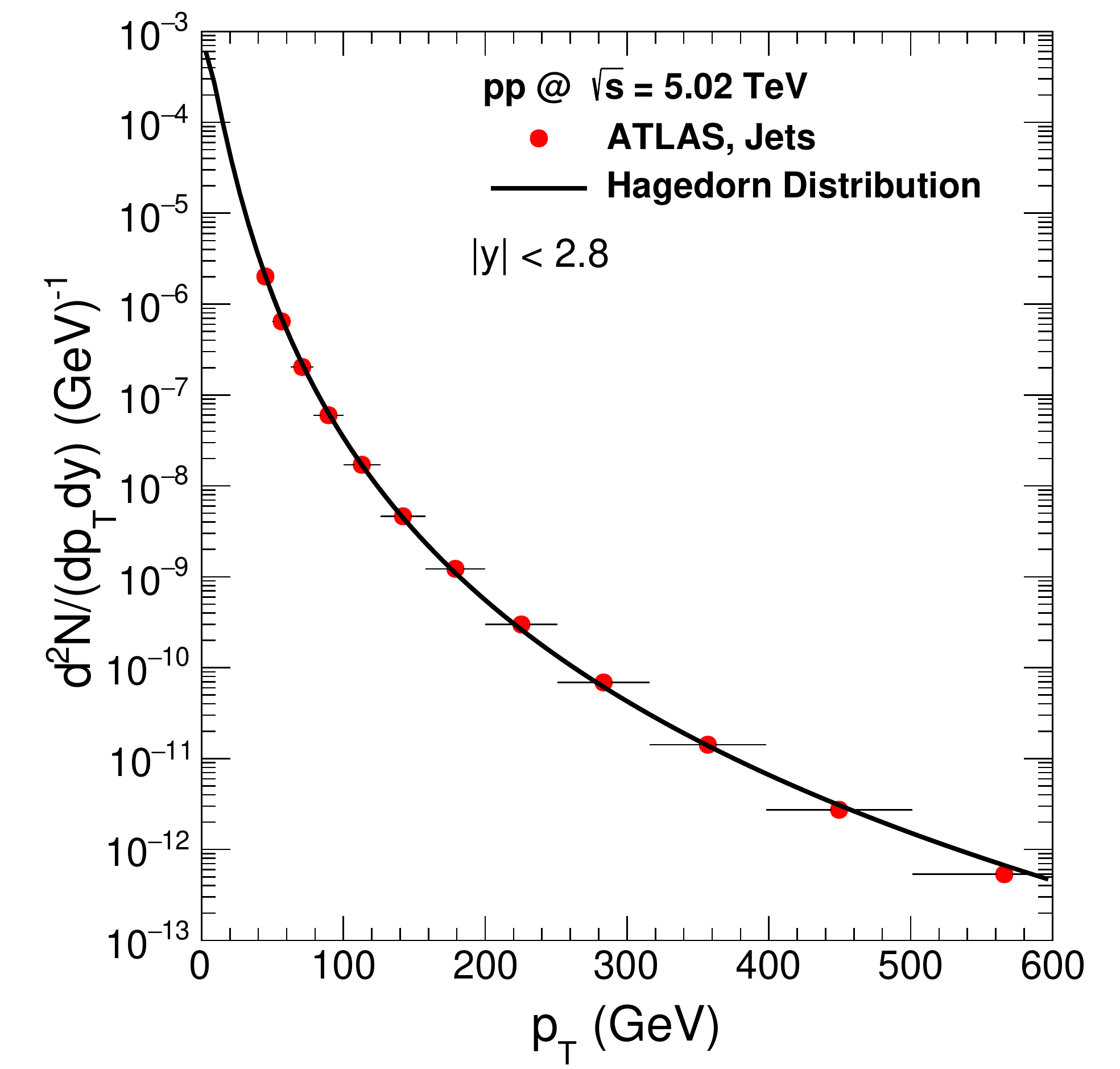}
\caption{The yields of the jets as a function of transverse 
momentum $p_{\rm{T}}$ for pp collision at $\sqrt{s}$ = 5.02 TeV measured by the 
ATLAS experiment \cite{Aaboud:2018twu}. The solid curve is the fitted Hagedorn 
distribution.}
\label{Figure11_jet_yield_pp_502tev}
\end{figure}

Figure~\ref{Figure11_jet_yield_pp_502tev} shows the yields of the jets
as a function of the transverse momentum $p_{\rm{T}}$ for pp collisions at $\sqrt{s}$
= 5.02 TeV measured by the ATLAS experiment \cite{Aaboud:2018twu}. The solid curve
is the Hagedorn distribution with the parameters given in 
Table~\ref{table0_charged_particles_jet_pT_spectra_tsallis_fitting_parameters_276_502_TeV}.

\begin{figure}[htp]
\centering
\includegraphics[width=0.85\linewidth]{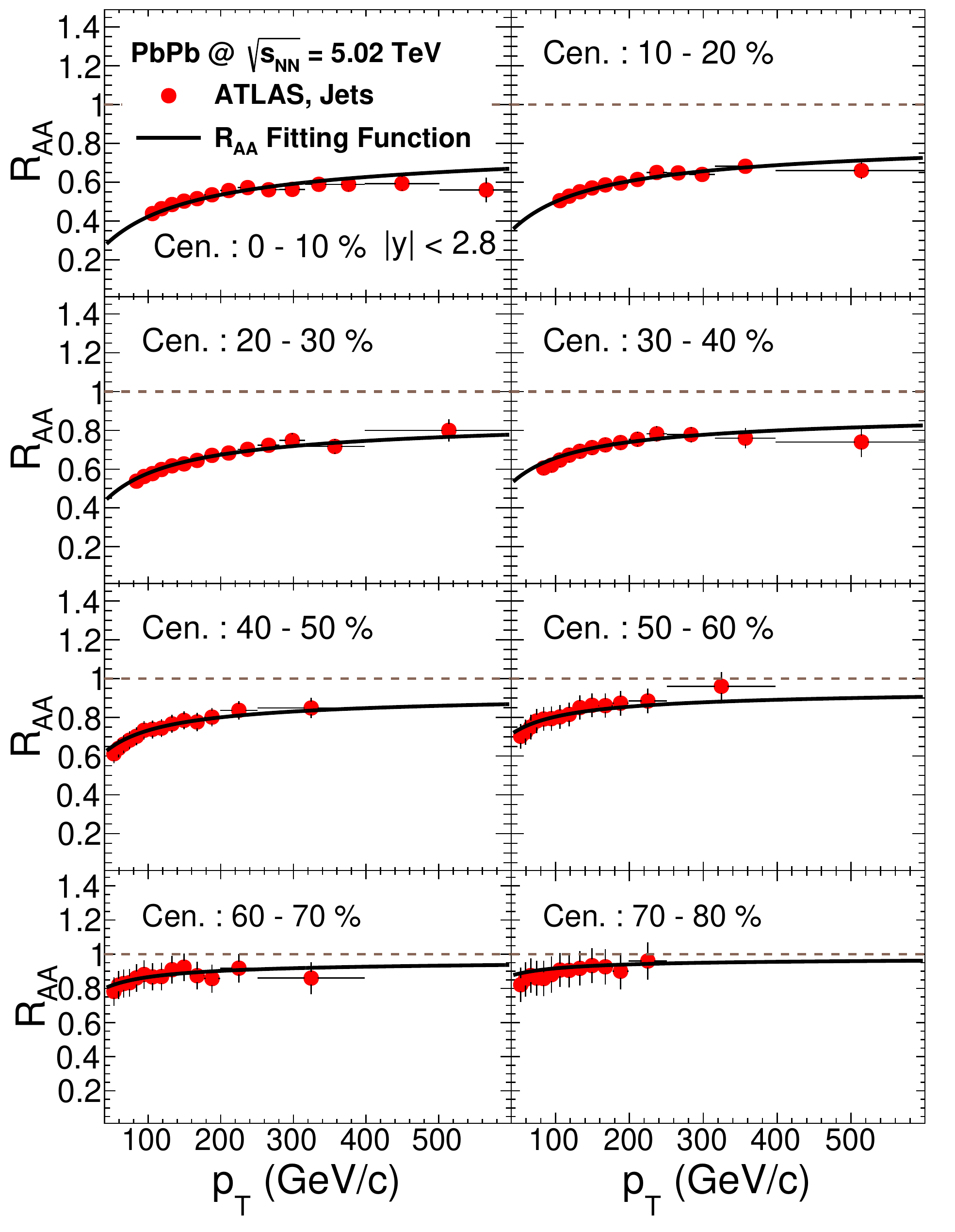}
\caption{The nuclear modification factor $R_{\rm{AA}}$  of the jets as a function
of transverse momentum $p_{\rm{T}}$ for various centrality classes in PbPb collisions at
$\sqrt{s_{\rm{NN}}}$ = 5.02 TeV measured by the ATLAS experiment \cite{Aaboud:2018twu}. 
The solid curves are the $R_{\rm{AA}}$ fitting function given by Eq.~\ref{nmf_raa_fitting_function}.}
\label{Figure12_Jet_particles_cms_RAA_spectra_com_fit_PbPb_502TeV}
\end{figure}

Figure~\ref{Figure12_Jet_particles_cms_RAA_spectra_com_fit_PbPb_502TeV} shows the
nuclear modification factor $R_{\rm{AA}}$ of the jets as a function of the
transverse momentum $p_{\rm{T}}$ for different centrality classes in PbPb collisions
at $\sqrt{s_{\rm{NN}}}$ = 5.02 TeV measured by the ATLAS experiment \cite{Aaboud:2018twu}.
 The solid curves are the $R_{\rm{AA}}$ fitting function (Eq.~\ref{nmf_raa_fitting_function}).
 The modeling of centrality dependence is done with $N_{\rm part}^\beta$ and the
value of exponent is obtained as $\beta=0.75$. 
  The extracted parameters of the energy loss
obtained by fitting the $R_{\rm{AA}}$ measured in different centrality classes of PbPb
collisions at $\sqrt{s_{\rm{NN}}}$ = 5.02 TeV are given in
Table~\ref{Table2_Jet_raa_fitting_parameter_PbPb_276_502_TeV},
along with the value of $\chi^{2}/\rm{NDF}$.
It shows that the $\Delta p_{\rm{T}}$ increases as $p_{\rm{T}}^{0.72}$
at all the values of $p_{\rm T}$ measured for jets similar to the case of jets at 
$\sqrt{s_{\rm{NN}}}$ = 5.02 TeV

\begin{figure}[htp]
\centering
\includegraphics[width=0.61\linewidth]{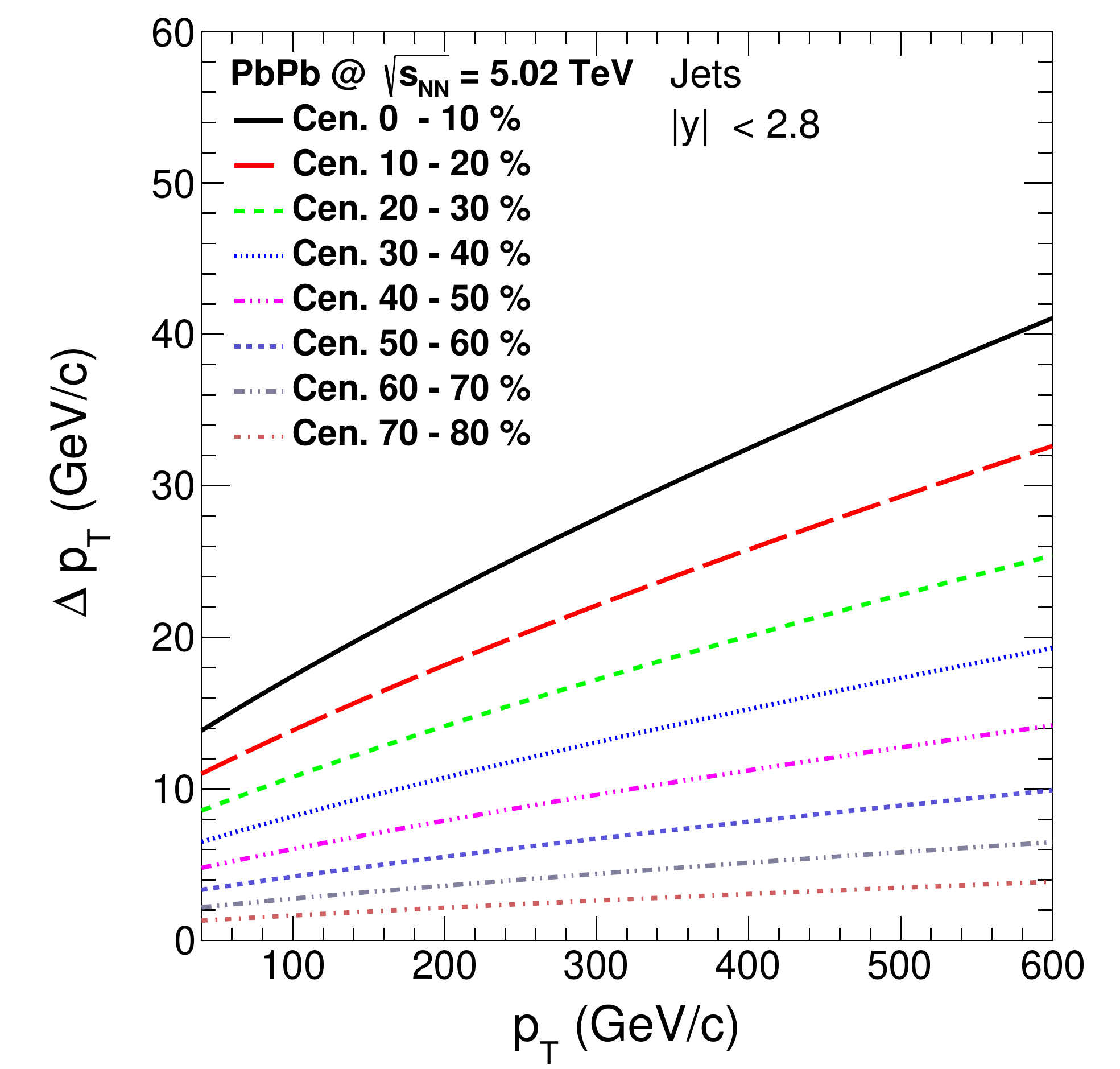}
\caption{The energy loss $\Delta p_{\rm{T}}$ of the jets as a function of transverse 
momentum $p_{\rm{T}}$ in PbPb collision at $\sqrt{s_{\rm{NN}}}$ = 5.02 TeV for different
centrality classes.}
\label{Figure13_Jet_particles_Del_pT_PbPb_502TeV}
\end{figure}

Figure~\ref{Figure13_Jet_particles_Del_pT_PbPb_502TeV} shows the energy loss
$\Delta p_{\rm{T}}$ of the jets as a function of the transverse momentum $p_{\rm{T}}$
for different centrality classes in PbPb collision at $\sqrt{s_{\rm{NN}}}$ = 5.02 TeV.
The $\Delta p_{\rm{T}}$ is obtained from  Eq.~\ref{Equation_Two} with the 
parameters given in Table~\ref{Table2_Jet_raa_fitting_parameter_PbPb_276_502_TeV}.
The $\Delta p_{\rm{T}}$ increases from peripheral to the most central collision regions.
The figure shows that the $\Delta p_{\rm{T}}$ increases almost linearly
at all the values of $p_{\rm T}$ measured for jets.

\begin{figure}[htp]
\centering
\includegraphics[width=0.61\linewidth]{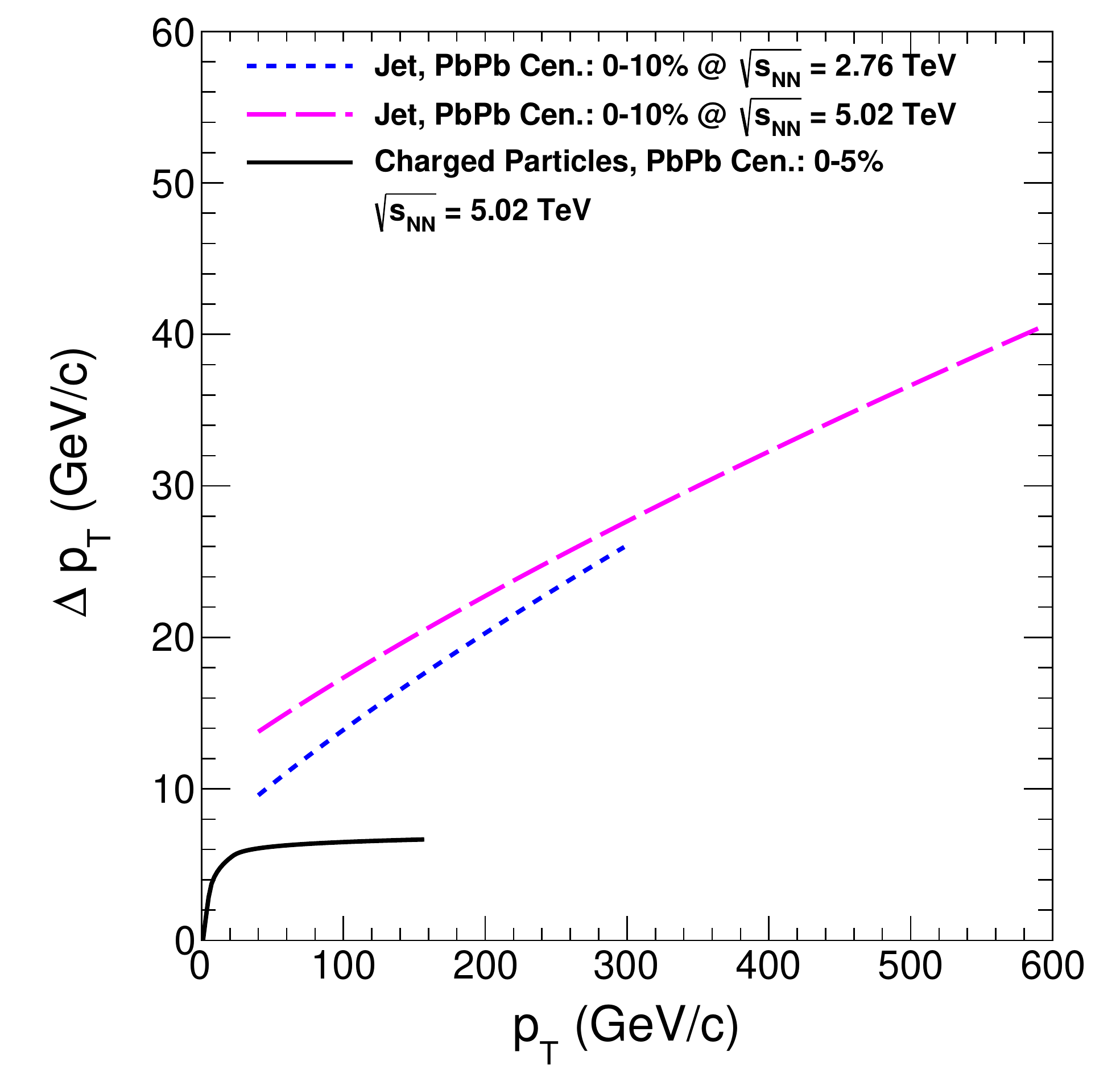}
\caption{The energy loss $\Delta p_{\rm{T}}$ of the jets as a function of transverse  
momentum $p_{\rm{T}}$ in the most central PbPb collision at $\sqrt{s_{\rm{NN}}}$ = 2.76 and 5.02 TeV. 
The $\Delta p_{\rm{T}}$ obtained for charged particles in the most central PbPb collision
at $\sqrt{s_{\rm{NN}}}$ = 5.02 TeV is also shown.}
\label{Figure14_jet_Del_pT_cen_0_10_PbPb_276_502TeV}
\end{figure}

Figure~\ref{Figure14_jet_Del_pT_cen_0_10_PbPb_276_502TeV} shows the energy loss
$\Delta p_{\rm{T}}$ of the jets as a function of the transverse momentum $p_{\rm{T}}$ 
in the most central (0-10\%) PbPb collision at $\sqrt{s_{\rm{NN}}}$ = 2.76 and 5.02 TeV.
These are compared with the $\Delta p_{\rm{T}}$ obtained for charged particles in
the 0-5\% centrality class of PbPb collision
at $\sqrt{s_{\rm{NN}}}$ = 5.02 TeV.
The energy loss $\Delta p_{\rm{T}}$ in case of jets for both the energies increases
with $p_{\rm{T}}$. The values of $\Delta p_{\rm{T}}$ for jets at 5.02 TeV is more than
that at 2.76 TeV. This behaviour at high $p_{\rm{T}}$ is very different from the
energy loss of charged particles which becomes  constant in these $p_{\rm{T}}$ regions.
The modeling of centrality dependence of energy loss has been done using
$N_{\rm part}^\beta$.
For charged particles, the centrality dependence of $p_T$ shift is found to
be $N_{\rm part}^{0.58}$ which corresponds to $L^{1.18}$.
The centrality dependence for jets at $\sqrt{s_{\rm NN}}$ = 2.76 TeV is found to be
$N_{\rm part}^{0.60}$. 
In case of jets at 5 TeV, the centrality dependence of energy loss is found to be
$N_{\rm part}^{0.75}$ corresponding to $L^{1.5}$ which means that the jets even at
very high energy are still away from complete coherent regime.

\clearpage

\section{Conclusions}

We presented a study of partonic energy loss with $p_T$ shift extracted from the measured
$R_{\rm{AA}}$ of charged particles and jets in PbPb collisions at $\sqrt{s_{\rm NN}}$ = 2.76
and 5.02 TeV in wide transverse momentum and centrality range.
  The functional form of energy loss given by
$\Delta p_{\rm T}$ has been assumed as power law with different power indices
in three different $p_{\rm T}$ regions driven by physics considerations.
The power indices and the boundaries of three $p_{\rm T}$ regions are obtained by
fitting the experimental data of $R_{\rm{AA}}$ as a function of $p_{\rm T}$ and centrality.
 The energy loss for light 
charged particles is found to increase linearly with $p_{\rm T}$ in low $p_{\rm T}$ region
below 5-6 GeV/$c$ and approaches a constant value in high $p_{\rm T}$ region above 25 GeV/$c$
with an intermediate power law connecting the two regions.
 The $\Delta p_{\rm{T}}$ at 5.02 TeV is similar but slightly more than
that at 2.76 TeV. 
In case of jets we consider only one $p_T$ region and it is found that for jets, the
energy loss increases almost linearly even at very
high $p_{\rm T}$.
 The modeling of centrality dependence of energy loss has been done using
$N_{\rm part}^\beta$.
 For charged particles, the centrality dependence of $p_T$ shift is found to
be $N_{\rm part}^{0.58}$ which corresponds to $L^{1.18}$.
 The centrality dependence for jets at $\sqrt{s_{\rm NN}}$ = 2.76 TeV goes as
$N_{\rm part}^{0.60}$. 
In case of jets at 5 TeV, the centrality dependence of energy loss is found to be 
$N_{\rm part}^{0.75}$ corresponding to $L^{1.5}$ which means that the jets even at
very high energy are still away from complete coherent regime.

\ \\

\noindent
{\bf References}


\end{document}